\newcommand{\AQO}[1]{\bar#1\gamma_3\gamma_5#1}
\newcommand{\PME}[1]{\langle\,p^\uparrow\,|\;#1\;|\,p^\uparrow\,\rangle}
\newcommand{\AV}{{\cal A}}
\newcommand{\M}{\hphantom{-}}
\newcommand{\Z}{\hphantom{0}}
\newcommand{\gev}{\,\mbox{GeV}}
\newcommand{\LQCD}{\Lambda_{\mbox{\scriptsize QCD}}}
\def\abbreviation#1#2#3{\def#1{#3 (#2)\def#1{#2}}}
\abbreviation {\BSR}   {BSR}  {Bjorken sum rule}
\abbreviation {\DIS}   {DIS}  {deep-inelastic scattering}
\abbreviation {\PQCD}  {PQCD} {perturbative QCD}
\abbreviation {\QPM}   {QPM}  {quark-parton model}
\documentstyle[12pt,world_sci]{article}
\let\thecase=\uppercase
\def\topline#1{\raisebox{48pt}[0pt][0pt]{\makebox[\textwidth]{#1}}\relax}
\begin{document}
\title{\topline{July 1994 \hfill MITH 94/11}
       \thecase{The Bjorken System of Equations}\\
       \thecase{and Nucleon Spin Structure-Function Data}
       \thanks{~Presented at DPF '94, Albuquerque, USA, August 1994}}
\author{\thecase{Philip G. Ratcliffe}\\
    \small\it Dip.\ di Fisica, Univ.\ di Milano\\
    \small\it via G. Celoria 16, 20133 Milano, Italy}
\maketitle
\begin{abstract}%
The status of the Bjorken sum rule is examined in the light of recent data
on the spin structure functions of the deuteron and proton obtained by the SMC
group, the neutron by the E142 group and the proton by the E143 group.
Combining the new data with that already obtained for the proton by the EMC
group and SLAC/YALE collaborations, we show that the Bjorken system of
equations is violated at the $2{-}3\sigma$ level. We also discuss in detail the
role of possible higher-twist contributions and higher-order PQCD corrections.
\end{abstract}

\noindent
Polarisation effects in general can provide valuable insight into the dynamics
of hadronic interactions and are extremely sensitive to the bound-state
structure, so elusive to theoretical approach\cite{PGR89}. In particular, the
\BSR\cite{Bj} is a measurable quantity that can be used to test theoretical
predictions. The experimental precision now attainable is at the ten-percent
level while, on the theoretical side, all relevant PQCD calculations have been
carried out to two-loop order\cite{Larin94} (i.e., one-percent level) and for
the \BSR\ itself to three loops\cite{Larin91}. Thus, one can consider such
comparisons as serious, indeed obligatory, tests of the applicability of PQCD
to such processes.

In the \QPM\ the structure function $g_1(x,Q^2)$\cite{Hughes93} is simply
related to polarised quark distributions, analogous to those for $F_1(x,Q^2)$:
\begin{equation}
 g_1(x,Q^2) = \mbox{$\frac{1}{2}$} \sum_f e_f^2 \, \Delta q_f(x,Q^2)
 \qquad\mbox{and}\qquad
 F_1(x,Q^2) = \mbox{$\frac{1}{2}$} \sum_f e_f^2 \,        q_f(x,Q^2).
\end{equation}
The quark densities are defined in the following manner:
\begin{equation}
 \Delta q_f(x,Q^2) = q_f^+(x,Q^2) - q_f^-(x,Q^2)
 \qquad\mbox{and}\qquad
        q_f(x,Q^2) = q_f^+(x,Q^2) + q_f^-(x,Q^2).
\end{equation}
where $q_f^\pm(x,Q^2)$ are the densities of quarks of flavour $f$ and positive
or negative helicity with respect to the parent hadron.

Experimentally one measures an asymmetry, the polarised structure function is
then extracted via
\begin{equation}
 g_1(x,Q^2) = \frac {A_1(x,Q^2) \, F_2(x,Q^2)} {2x\,(1+R(x,Q^2))} ,
\end{equation}
where $R_1(x,Q^2)$ is the ratio of longitudinal to transverse unpolarised
structure functions and $A_1(x,Q^2)$ is the measured asymmetry.

The Bjorken sum rule\cite{Bj} with PQCD corrections reads
\begin{equation}
 \Gamma_1^{p-n} = \int_0^1 dx\,g_1^{p-n}(x,Q^2) = \mbox{$\frac{1}{6}$}\,g_A\,
 \left[1-\alpha_s/\pi-c_2(\alpha_s/\pi)^2-\dots\right],
\end{equation}
the coefficients being known to third order\cite{Larin91}.

The full SU(3) algebra of the baryon octet admits three independent quantities,
which may be expressed in terms of the SU(3) axial-vector couplings:
\begin{equation} \label{BJsystem}
\begin{array}{c@{\;=\;}c@{\;=\;}l}
 \PME{\AQO{u} - \AQO{d}}            & \PME{\AV_3} &        g_A , \\
 \PME{\AQO{u} + \AQO{d} - 2\AQO{s}} & \PME{\AV_8} & \tilde g_A , \\
 \PME{\AQO{u} + \AQO{d} + \AQO{s}}  & \PME{\AV_0} &        g_0 .
\end{array}
\end{equation}
The right-hand sides of the first two equations correspond to measured
constants ($g_A{=}1.2573{\pm}0.0028$\cite{PDG94} and
$\tilde{g}_A{=}0.629{\pm}0.039$\cite{PGR90}), but the third ($g_0$),
corresponding to the flavour-singlet axial-vector current, is unknown. Thus a
direct prediction for, say, just the proton integral is not possible.
A further combination of the $u$, $d$ and $s$ axial-current matrix elements is
accessible in $\nu$-$p$ elastic scattering\cite{Ahre87} and thus would allow an
exact prediction for single nucleon targets. Unfortunately, the precision of
such measurements is still very poor.

However, good arguments can be made for setting the strange-quark matrix
element equal to zero\cite{Ellis74}: there are few strange quarks in the proton
and they are concentrated below $x_B{\simeq}0.1$, where all correlations are
expected to die out. Thus, the last two matrix elements of
eqs.~(\ref{BJsystem}) might be expected to be equal, leaving only two
independent quantities and allowing predictions for the proton and neutron
separately:
\begin{equation}
 \Gamma_1^{p(n)} = (-)\mbox{$\frac{1}{12}$}        g_A
                    + \mbox{$\frac{5}{36}$} \tilde g_A
                    + \mbox{$\frac{1}{3}$}  \PME{\AQO{s}} ,
\end{equation}
where the last term is then assumed negligible. For clarity, the PQCD
corrections have been suppressed.
Conversely, these equations may be used to extract the
value of either the strange-quark or singlet axial-vector matrix element, given
the value of $\Gamma_1$.
There is no space here to discuss the problem of the strange-quark spin; the
interested reader is referred to\cite{Prep88,Prep90a}, where a bound on
the non-diffractive component and thus on the strange-quark polarisation was
derived. The result of this analysis is the following bound:
$\left|\int\Delta{s}\right|{\le}0.02$.

We now compare the results obtained by the three experiments with theoretical
predictions based on the above.
In performing the calculations we have used the very precise value of
$\LQCD^{(4)}$ recently extracted in a three-loop analysis of scaling violations
in \DIS\cite{Pare94}, which is thus most suitable for our purposes. This
analysis also allows an examination of the improvement obtained on increasing
the order of the perturbation theory analysis. Let us take the opportunity to
stress that for any analysis to be consistent, all quantities involved must be
evaluated at the same loop order and that, in particular, it is meaningless to
insert a two-loop $\alpha_s$ into a three-loop expression.
\begin{eqnarray}
\makebox[2cm][l]{EMC\cite{EMC88}}  \Gamma_1^p(11\gev^2)
 &=& \M 0.126 \pm 0.010 \pm 0.015 \nonumber \\
\makebox[2cm][l]{SMC\cite{SMC94}}  \Gamma_1^p(10\gev^2)
 &=& \M 0.136 \pm 0.011 \pm 0.011 \nonumber \\
\makebox[2cm][l]{E143\cite{E143}}  \Gamma_1^p(3\gev^2)\Z
 &=& \M 0.133 \pm 0.004 \pm 0.012           \\
\makebox[2cm][l]{SMC\cite{SMC93}}  \Gamma_1^d(5\gev^2)\Z
 &=& \M 0.023 \pm 0.020 \pm 0.015 \nonumber \\
\makebox[2cm][l]{E142\cite{E142}}  \Gamma_1^n(2\gev^2)\Z
 &=&  - 0.022 \pm 0.006 \pm 0.009 \nonumber \\
[6pt]
 \Gamma_1^p(11\gev^2)
 &=& \M 0.182 \pm 0.006 + \mbox{$\frac13\int$}\Delta{s} \nonumber \\
 \Gamma_1^p(10\gev^2)
 &=& \M 0.182 \pm 0.006 + \mbox{$\frac13\int$}\Delta{s} \nonumber \\
\makebox[0cm][l]{\raisebox{ 8pt}[0pt][0pt]{Ellis-}}%
\makebox[2cm][l]{\raisebox{-8pt}[0pt][0pt]{Jaffe }}
 \Gamma_1^p(3\gev^2)\Z
 &=& \M 0.179 \pm 0.006 + \mbox{$\frac13\int$}\Delta{s}           \\
 \Gamma_1^d(5\gev^2)\Z
 &=& \M 0.085 \pm 0.006 + \mbox{$\frac13\int$}\Delta{s} \nonumber \\
 \Gamma_1^n(2\gev^2)\Z
 &=&  - 0.010 \pm 0.006 + \mbox{$\frac13\int$}\Delta{s} \nonumber
\end{eqnarray}
The short-fall in the proton measurements with respect to the Ellis-Jaffe
prediction (taking ${\int}\Delta{s}{=}0$) is immediately obvious. This
observation led to the coining of the phrase {\em Spin Crisis\/}. A similar
(though less striking) observation may be made for the SMC deuteron integral.
In contrast, the neutron sum rule appears well satisfied by the E142 data. In
terms of the strange-quark contribution, both the EMC and SMC measurements
imply ${\int}\Delta{s}{\simeq}{-}0.15$ while that of E142 leads to
${\int}\Delta{s}{\simeq}{-}0.04$.

A measure of the discrepancy between the data and theory may be obtained by
extracting the singlet axial-vector matrix element: the results are
\begin{equation}
\begin{array}{r@{\;=\;}l}
\Delta q
 & 0.14 \pm 0.16 \qquad\makebox[3.5cm][l]{EMC proton}    \\
 & 0.21 \pm 0.14 \qquad\makebox[3.5cm][l]{SMC proton}    \\
 & 0.21 \pm 0.12 \qquad\makebox[3.5cm][l]{E143 proton}   \\
 & 0.06 \pm 0.22 \qquad\makebox[3.5cm][l]{SMC deuteron}  \\
 & 0.20 \pm 0.08 \qquad\makebox[3.5cm][l]{global proton} \\
 & 0.51 \pm 0.10 \qquad\makebox[3.5cm][l]{E142 neutron,}
\end{array}
\end{equation}
where $\Delta{q}$ is the sum of quark polarisations as in eqs.~\ref{BJsystem}.
Alternatively, one can fit for the strange-quark spin
contribution\cite{Prep93b}. Taking the SLAC proton and neutron data and
performing completely consistent fits at one- two- and three-loop order we
obtain respectively $\chi^2{=}3.7$, 3.8 and 3.2 for one degree of freedom.
Using the Particle Data Group\cite{PDG94}  preferred value of
$\LQCD^{(4)}{=}260^{+56}_{-46}$ in a two-loop fit (for consistency with the
extraction of $\Lambda$), the situation is marginally improved to give
$\chi^2{=}2.8$.

Given the low $Q^2$ of the SLAC data, one should naturally worry about the
possibility of higher-twist ``contamination''. There are two approaches to this
problem: either one attempts to estimate theoretically the size of such effects
(e.g., using a bag model\cite{Ji93} or QCD sum rules\cite{Ross93}) or one
deduces limits from the well-documented higher-twist behaviour of the
unpolarised data\cite{PGR93}. In either case it turns out that the magnitude of
higher-twist contributions to \DIS\ is far too small to have any real impact,
even on the SLAC neutron data (by a strange quirk, the higher-twist
contribution to $g_1^n$ is typically much smaller even than that in the case of
$g_1^p$).

It is interesting to ask what occurs if the normalization condition on the
Wilson coefficients is relaxed, i.e., if one ignores PQCD and uses current
algebra only to fix ratios of matrix elements\cite{PGR93}. In this case, using
our strange-quark bound to effectively set $\Delta{s}{=}0$, any one data set
may be used to fix the overall normalization. The EMC proton data, for example,
then lead to the following ``prediction'' for the neutron:
$0.002{\le}\Gamma_1^n{\le}-0.026$, in rather good agreement with the SLAC data.
Alternatively, the quark spins may be deduced from the proton and neutron data:
one arrives at the following relation:
\begin{equation}
\Gamma_1^n = -\mbox{$\frac1{11}$}\Gamma_1^p + \mbox{$\frac23$}\Delta s,
\end{equation}
which leads to $\Delta{s}=-0.03\pm0.03$, again perfectly compatible with our
bound.



\begin{thebibliography}{00.}

\bibitem{PGR89}
P.G.~Ratcliffe, in {\em Problems of Fundamental Modern Physics\/} (World
Scientific 1990, eds.\ R.~Che\-rubini, P.~Dalpiaz and B.~Minetti), p.~233.

\bibitem{Bj}
J.D.~Bjorken, {\it Phys.\ Rev.\/} {\bf 148} (1966)~1467; {\bf D1} (1970)~1376.

\bibitem{Larin94}
S.A.~Larin, CERN preprint CERN-TH.7208/94.

\bibitem{Larin91}
S.A.~Larin and J.A.M.~Vermarseren, {\it Phys.\ Lett.\/} {\bf B259} (1991)~345.

\bibitem{Hughes93}
A review of the experimental and theoretical situation can be found in:
E.~Hughes, {\em Polarized Lepton-Nucleon Scattering}, presented at the 21st
Annual SLAC Summer Institute (July 1993).

\bibitem{PDG94}
PDG, M.~Anguilar-Benitez {\it et al.\/}, {\it Phys.\ Rev.\/} {\bf D50}~(1994).

\bibitem{PGR90}
P.G.~Ratcliffe, {\it Phys.\ Lett.\/} {\bf B242} (1990)~271.

\bibitem{Ahre87}
L.A.~Ahrens {\it et al.\/}, {\it Phys.\ Rev.\/} {\bf D35} (1987)~785.

\bibitem{Ellis74}
J.~Ellis and R.L.~Jaffe, {\it Phys.\ Rev.\/} {\bf D9} (1974)~1444;
{\it ibid.\/} {\bf D10} (1974)~1669.

\bibitem{Prep88}
G.~Preparata, and J.~Soffer, {\it Phys.\ Rev.\ Lett.\/} {\bf 61} (1988)~1167;
{\it Erratum\/} {\bf 62} (1989)~1213.

\bibitem{Prep90a}
G.~Preparata, P.G.~Ratcliffe and J.~Soffer,
{\it Phys.\ Rev.\/} {\bf D42} (1990)~930;
{\it Phys.\ Lett.\/} {\bf B273} (1991)~306.

\bibitem{Pare94}
G.~Parente, A.V.~Kotikov and V.G.~Krivokhizhin, Irvine preprint UCI-TR/94-4.

\bibitem{EMC88}
EMC, J.~Ashman {\it et al.\/}, {\it Phys.\ Lett.\/} {\bf B206} (1988)~364;
{\bf B328} (1989)~1.

\bibitem{SMC94}
SMC, D.~Adams {\it et al.\/}, {\it Phys.\ Lett.\/} {\bf B329} (1994)~399.

\bibitem{E143}
E143, R.~Arnold {\it et al.\/}, presented at the Conf. on the Intersections of
Particle and Nuclear Physics (St.~Petersburg Florida, 1994).

\bibitem{SMC93}
SMC, B.~Adeva {\it et al.\/}, {\it Phys.\ Lett.\/} {\bf B302} (1993)~533.

\bibitem{E142}
E142, P.L.~Anthony {\it et al.\/},
{\it Phys.\ Rev.\ Lett.\/} {\bf 71} (1993)~959.

\bibitem{Prep93b}
G.~Preparata and P.G.~Ratcliffe, Milano preprints MITH~93/9, MITH~93/12 and
MITH~93/15.

\bibitem{Ji93}
X.~Ji and P.~Unrau, MIT preprint MIT-CTP-2232.

\bibitem{Ross93}
G.G.~Ross and R.G.~Roberts, Rutherford lab.\ preprint RAL-93-092.

\bibitem{PGR93}
P.G.~Ratcliffe, Milano preprint MITH~93/28.

\end{thebibliography}
\end{document}